\documentclass[prb,aps,twocolumn,groupedaddress,floatfix,citeautoscript,showpacs]{revtex4}
\usepackage{graphicx,rotating,subfigure,amsmath,amsfonts,amssymb,delarray,color}
\usepackage{dsfont}
\usepackage[T1]{fontenc}

\def\l{\left}
\def\r{\right}
\def\12{\frac{1}{2}}

\newcommand{\be}{\begin{equation}}
\newcommand{\ee}{\end{equation}}
\newcommand{\bea}{\begin{eqnarray}}
\newcommand{\eea}{\end{eqnarray}}

\graphicspath{{figs/}}

\begin{document}
\bibliographystyle{apsrev} 

\title{Propagation of a single hole defect in the one-dimensional Bose-Hubbard model}

\author{F. Andraschko}
\author{J. Sirker}
\affiliation{Department of Physics and
  Astronomy, University of Manitoba, Winnipeg R3T 2N2, Canada}
\affiliation{Department of Physics and Research Center OPTIMAS,
  Technical University Kaiserslautern,
  D-67663 Kaiserslautern, Germany}

\date{\today}

\begin{abstract}
  We study nonequilibrium dynamics in the one-dimensional Bose-Hubbard
  model starting from an initial product state with one boson per site
  and a single hole defect. We find that for parameters close to the
  quantum critical point, the hole splits into a core showing a very
  slow diffusive dynamics, and a fast mode which propagates
  ballistically.  Using an effective fermionic model at large Hubbard
  interactions $U$, we show that the ballistic mode is a consequence of
  an interference between slow holon and fast doublon dynamics, which
  occurs once the hole defect starts propagating into the bosonic
  background at unit filling. Within this model, the signal velocity
  of the fast ballistic mode is given by the maximum slope of the
  dispersion of the doublon quasiparticle in good agreement with the
  numerical data. Furthermore, we contrast this global quench with the
  dynamics of a single hole defect in the ground state of the
  Bose-Hubbard model and show that the dynamics in the latter case is
  very different even for large values of $U$. This can also be seen
  in the entanglement entropy of the time evolved states which grows
  much more rapidly in time in the global quench case.
\end{abstract}


\maketitle

\section{Introduction}

An investigation of the states of matter and excitations of a quantum
system is usually based on its dynamical response to external
perturbations. This includes frequency-dependent response functions
and scattering cross sections tested in condensed matter
experiments,\cite{kittelbook,ashcroftbook} as well as time-dependent
correlation functions measured in experiments on cold atomic
gases.\cite{greiner2002,kinoshita2006,cheneau2012,fukuhara2013nat,fukuhara2013nph,hild2014,arxivxia,bloch2008}
The investigation of time-dependent correlation functions in and out
of equilibrium has recently attracted renewed interest due to
experiments on cold atomic gases in optical lattices which provide
useful schemes for the preparation of quantum states and the tuning of
interactions.\cite{bloch2008,paredes2004,kinoshita2004} Recent
developments allow, in particular, for single-site addressability,
both in preparation and detection\cite{gericke2008, bakr2009,
  wuertz2009, sherson2010, endres2011}.

A lot of attention has been focussed on the dynamics of quasi
one-dimensional systems where quantum fluctuations are strong and can
prevent the ground state from attaining any type of long-range order.
In this case, the system can often be described by Luttinger liquid
(LL) theory \cite{giamarchibook}, a hydrodynamic approach for the
collective excitations (density fluctuations), which propagate with an
interaction-dependent velocity.  Recent studies, however, have shown
that LL theory is not sufficient to fully describe the dynamics of
such systems even at zero temperature. Nonlinearities of the
dispersion have to be taken into account, which lead to x-ray
edge-type singularities in response functions and new quasiparticles.
An extension of standard LL theory, the so-called nonlinear Luttinger
liquid (NLL) theory, has been developed, which is capable of describing
the collective excitations and the new quasiparticles on an equal
footing.\cite{imambekov2012} This allows, for example, to calculate
the dynamical structure factor for the Lieb-Liniger
model\cite{liebliniger1963}---the continuum limit of the Bose-Hubbard
model in the superfluid phase--- which shows power-law singularities
at the edges of its support.\cite{imambekov2012} Very recently, it has
been demonstrated that the NLL approach can also be extended to small
finite temperatures \cite{arxivkarrasch}.  One of the main findings of
the latter study is that the combination of quantum and thermal
fluctuations can dramatically alter the dynamical response compared to
the zero temperature case. For the spin-1/2 XXZ chain, in particular,
the autocorrelation function at finite temperature
and long times was shown to exhibit a slow diffusive decay $\sim 1/\sqrt{t}$ in time
$t$, explaining the diffusive response measured in NMR and $\mu$SR
experiments \cite{sirker2009,sirker2011,thurber2001}. To also
understand spin and energy transport in the linear response regime,
one has to identify, in addition, all the local and quasilocal
conserved charges of the {\it microscopic model} under consideration.
For the XXZ chain, for example, the spin current is only partly
conserved, so that a ballistic contribution coexists with a diffusive
contribution.\cite{sirker2009,sirker2011} This result shows that even
the linear response in a very simple strongly correlated
one-dimensional quantum model can already be quite complicated and
very different from phenomenological theories \cite{degennes1958}.

Dynamics and transport far from equilibrium are, on the other hand,
much less understood so far even for the simplest interacting models.
Much of the interest has concentrated on quantum quenches, i.~e., the
dynamics ensuing after preparing the system in the ground state of a
Hamiltonian, and then suddenly changing one of its microscopic
parameters. The expectation values of local observables at long times
after the quench in a generic, infinitely large quantum system are
expected to become time independent and to be described by a Gibbs
ensemble.\cite{rigol2008} For integrable quantum systems, on the other
hand, the additional local and quasilocal conserved charges have to be
taken into account as well, leading to a generalized Gibbs ensemble
\cite{rigol2007,fagotti2013,sirker2014}.  While (generalized) Gibbs
ensembles can help us to calculate the equilibrated values of local
observables after a quantum quench for a system in the thermodynamic
limit, they are not helpful in identifying the dominant relaxation
processes and in describing the nonequilibrium dynamics itself. A
notable known feature in the quench dynamics of a model with
short-range interactions, starting from an initial state with a finite
correlation length, is the existence of Lieb-Robinson bounds which
state that correlations can only spread with a finite
speed.\cite{lieb1972,bravyi2006} The emergence of an effective light
cone has also been observed in
experiments.\cite{cheneau2012,langen2013} Furthermore, such bounds can
exist for finite-temperature initial states as well.\cite{bonnes2014}

A possible way to systematically probe nonequilibrium dynamics and
transport beyond linear response is to consider local
perturbations.\cite{kollath2005,kleine2008,langer2009,langer2011,fukuhara2013nph,fukuhara2013nat,karrasch2014}
The advantage of this approach is that local perturbations can easily
be created in an experimental setup of cold atoms, and that one can
follow their evolution in time by measuring one-point functions
instead of two-point correlators.  For the Bose-Hubbard model, for
example, the time evolution of local Gaussian density perturbations on
top of the ground state has been studied by numerical density matrix
renormalization group (DMRG) methods.\cite{kollath2005} In this work,
it has been shown that for weak and intermediate interactions, small
density perturbations travel through the system at a velocity which is
consistent with the sound velocity in the corresponding continuum, exactly
solvable Lieb-Liniger model. The expansion of
density-wave packets in the one-dimensional Bose-Hubbard model has
also been studied experimentally in an out-of-equilibrium
situation.\cite{ronzheimer2013} Here, a gas of $^{39}$K atoms was
prepared in a harmonic confinement. The confinement was then lowered
so that the atoms started to expand into the empty optical lattice. A
slowing down of the expansion of the cloud was observed near the
critical point in the ground state phase diagram.
The same experiment in two dimensions showed, in addition, a
separation of the cloud into a diffusive core and a background which
spreads ballistically. An important feature of such expansion dynamics
is that more than one velocity can be present due to the formation of
bound states. Examples for such behavior are the {\it XXZ} model at
large repulsive nearest-neighbor
interaction\cite{ganahl2012,fukuhara2013nat,alba2013,liu2014}, as well
as the Bose-Hubbard model, where the onsite repulsive interaction can
lead to effectively bound states of
doublons.\cite{jreissaty2013,degliespostiboschi2014}

In this paper, we will study the non-equilibrium dynamics of one of
the simplest local perturbations: a single hole defect in an initial
product state with one boson per lattice site under the unitary time
evolution of a Bose-Hubbard Hamiltonian. This setup can easily be
realized using cold atomic gases, which offer single-site
addressability in the preparation of initial states and in the
detection of bosons after a quantum quench.\cite{gericke2008,
  bakr2009, wuertz2009, sherson2010, endres2011} Here, the hole defect
spreads into a background which itself displays nontrivial dynamics,
contrasting our setup from the case of an expansion into an empty
lattice, which has been studied thoroughly for the
bosonic\cite{rodriguez2006,jreissaty2011,muth2012,ronzheimer2013,jreissaty2013,vidmar2013,degliespostiboschi2014}
and
fermionic\cite{heidrichmeisner2009,kajala2011,karlsson2011,schneider2012,langer2012}
versions of the Hubbard model. We will show, in particular, that the
defect will either spread ballistically through the lattice or
separate into a diffusive core coexisting with a fast ballistic mode,
depending on the strength of the Hubbard interaction $U$. This
phenomenon is reminiscent of the coexistence of ballistic and
diffusive transport in the XXZ model at finite temperature, however,
here it is not related to local or quasilocal conserved charges, but
is rather, as we will show, a consequence of an interference effect
between slow holon and fast doublon quasiparticle dynamics.

Our paper is organized as follows: In Sec.~\ref{Model} we define the
model, the initial states, and the time-dependent correlation
functions which we study in the following. In Sec.~\ref{Method} we
explain the numerical method used to simulate the quench dynamics and
introduce an effective fermionic model\cite{barmettler2012} to
approximate the dynamics for large interactions. In Sec.~\ref{Results}
we present the results of the numerical calculations and analyze these
results using the effective fermionic model. In Sec.~\ref{Conclusions}
we summarize our main results and discuss some of the remaining open
questions.

\section{Model}
\label{Model}
We consider the one-dimensional Bose-Hubbard model, 
\begin{align}
\label{eq:hbhm}
H&= J\sum_j (b_j^\dagger b_{j+1}+h.c.) + \frac{U}{2}\sum_j n_j(n_j-1),
\end{align}
where $b_j^\dagger$ ($b_j$) creates (annihilates) a boson at site $j$,
$n_j=b_j^\dagger b_j$ is the local density operator, $U$ is the strength
of the on-site repulsive interaction, and in the
following, we set $J=1$. At unit filling, this model exhibits a
quantum phase transition at $U_c\approx 3.37$ between a superfluid and
a Mott insulating state.\cite{kuehner1998} We are interested in the
time-dependent density profile $C_r(j,t)$ after removing a single
particle at the center of the chain from a prepared uniform initial
state $|\Psi_r\rangle$
\begin{align}
\label{eq:corrfcts}
C_r(j,t) &= \langle \Psi_r|b_{j=0}^\dagger e^{iHt} n_j e^{-iHt}
b_{j=0}|\Psi_r\rangle.
\end{align}
Our main focus is on the time evolution starting from the initial
product state $|\Psi_{r\equiv p}\rangle\equiv
\prod_jb_j^\dagger|0\rangle$. To understand the unique features of the
non-equilibrium dynamics, we will compare our results with the
equilibrium correlation function obtained by using the ground state
$|\Psi_{r\equiv g}\rangle$ of the time-evolving Hamiltonian $H$ at
unit filling as the initial state in Eq.~\eqref{eq:corrfcts}.
Alternatively, the latter case can be viewed as a local
quench---defined as a local modification of the Hamiltonian or the
initial state---while in the former case, an additional global quench
from the ground state at infinite repulsion is performed.

\section{Method}
\label{Method}
\subsection{Numerics}
We calculate the correlation function \eqref{eq:corrfcts} using the
light cone renormalization group (LCRG) algorithm.\cite{enss2012} This
algorithm is based on the time-dependent DMRG\cite{white1992,
  vidal2004, white2004, daley2004} and yields results directly in the
thermodynamic limit. In our simulations we restrict the maximum number
of bosons per lattice site to $5$ for $U\le 6$, and to $3$ for $U> 6$.
By comparing with simulations where the maximum number of bosons is
smaller we make sure that the results are converged on the scale of
the figures we present in the following. Furthermore, we make sure
that the discarded weight in each renormalization group step is
smaller than $10^{-8}$. For the simulation times shown, this requires
to increase the number of states kept in the truncated Hilbert space
up to $8\,000$.

\begin{figure}[ht!]
  \centering
  \includegraphics[width=\columnwidth]{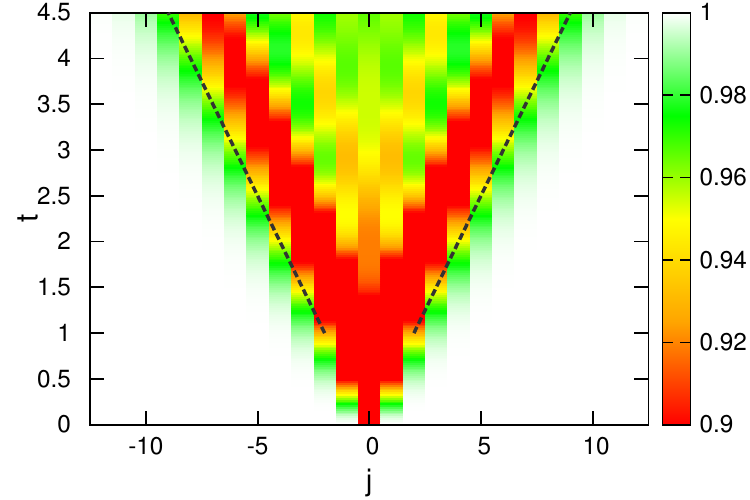}
  \caption{(Color online) The density profile $C_p(j,t)=1-\mathcal{J}_j^2(2t)$ for $U=0$
    and $U=\infty$. The dashed lines show the linear spreading of the defect density, $j=\pm v t$, with a defect velocity $v=2$.
  }
  \label{fig:bessel}
\end{figure}
For the calculations where the initial state is the ground state
$|\Psi_g\rangle$, we first perform an imaginary time evolution with
the correct chemical potential $\mu$ to project an arbitrary starting
vector onto the ground state at unit filling. The chemical potential
is required because a canonical simulation in the thermodynamic limit
is impossible. We check that $|\langle
\Psi_g|n_j|\Psi_g\rangle-1|<10^{-6}$, i.~e. we ensure convergence in
the density to more than 6 digits.

\subsection{Analytical results and approximations}
\label{ss:analytics}
The Bose-Hubbard model, Eq.~\eqref{eq:hbhm}, has two integrable
limits: the limit of free bosons, $U=0$, and the limit of hard-core
bosons, $U=\infty$. In both cases, we are dealing with single-particle
dynamics and the density profile can be calculated exactly:
\begin{equation}
\label{free_case}
C_p^{U=0}(j,t)=C_p^{U=\infty}(j,t)=1-\mathcal{J}_j^2(2t),
\end{equation}
where $\mathcal{J}_j$ is the $j$th Bessel function of the first kind.
The density profile for these cases is shown in Fig.~\ref{fig:bessel}:
The defect density spreads in a light cone with a velocity $v=2$.  If
we create a doublon instead of a hole defect, i.e., interchange
$b_{j=0}^\dagger \leftrightarrow b_{j=0}$ in Eq.~\eqref{eq:corrfcts},
the corresponding correlation function $\widetilde C_p$ is
given by $\widetilde C_p^{U=0}(j,t)=1+\mathcal{J}_j^2(2t)$ and
$\widetilde C_p^{U=\infty}(j,t)=1+\mathcal{J}_j^2(4t)$. In this case,
the velocity of the expansion is given by $v=2$ for $U=0$ and $v=4$
for $U=\infty$. The by a factor of $2$ larger velocity in the latter case
is a consequence of the Bose factor.

In the following, we will call any dynamics similar to the one shown
in Fig.~\ref{fig:bessel} ballistic. For a more precise definition, we
follow Refs.~$ $\onlinecite{langer2009, vidmar2013} and consider
\begin{align}
\label{eq:rsq}
R^2(t)&= \sum_{j=-\infty}^{\infty}j^2 (1-C_p(j,t)),
\end{align}
where $R(t)$ is a weighted measure for the distance over which the
defect density has spread at time $t$. For the free and hardcore boson
cases, the sum can be calculated exactly, using Eq.~\eqref{free_case}
and a standard identity for Bessel functions, resulting in
\begin{align}
R^2(t)&= \sum_{j=-\infty}^\infty j^2\mathcal{J}_j^2(2t)=2t^2.
\end{align}
The functional dependence of $R^2$ on time is thus the same as for
a linearly dispersing, ballistic particle,
\begin{align}
R^2(t)&= \sum_{j=-\infty}^\infty \frac{j^2}{2}\left[\delta(j+vt)+\delta(j-vt)\right]=v^2t^2. 
\end{align}

For large but finite $U$, an exact solution is not possible. Since high
occupancies are suppressed, we can, however, approximate the dynamics
in this limit by restricting the local Hilbert space to three states
only: $|0\rangle_j$, $|1\rangle_j$, and $|2\rangle_j$.  We can now
interpret the state $|1\rangle_j$ as the vacuum, the doublon state
$|2\rangle_j$ as a spin-up particle, and the holon state $|0\rangle_j$
as a spin-down particle. This is made explicit by the mapping
\begin{equation}
\label{hcbosons}
b_j=\sqrt{2}\,\widetilde b_{j\uparrow} + \widetilde b_{j\downarrow}^\dagger
\end{equation}
where $\widetilde b_{j\sigma}$ are hardcore bosons.\cite{barmettler2012}
The hardcore constraint for identical particles can be fulfilled by a
second mapping onto fermions
\begin{align}
\label{eq:bosefermi}
\widetilde b_{j\uparrow}=Z_jc_{j\uparrow}\quad , \quad \widetilde b_{j\downarrow}=Z_j e^{i\pi n_{j\uparrow}} c_{j\downarrow}^\dagger,
\end{align}
where $Z_j=\prod_{j'<j}e^{i\pi \sum_{\sigma}n_{j\sigma}}$ is a
Jordan-Wigner string, $c_{j\sigma}$ are fermionic operators, and
$n_{j\sigma}=c^\dagger_{j\sigma}c_{j\sigma}$. Applying both
transformations maps the bosonic Hamiltonian \eqref{eq:hbhm} in the
restricted Hilbert space onto the fermionic Hamiltonian in momentum
space,
\begin{eqnarray}
\label{eq:HUF}
H &=& \sum_{k\sigma}E_{k\sigma} c_{k\sigma}^\dagger c_{k\sigma} +i \sum_{k}\Delta_k(c_{k\uparrow}^\dagger c_{\bar{k}\downarrow}^\dagger - c_{\bar{k}\downarrow}c_{k\uparrow}) \nonumber \\
&+&\frac{V}{N}\sum_{k,k',q} c^\dagger_{k\uparrow}c_{k-q\uparrow}c^\dagger_{k'\downarrow}c_{k'+q\downarrow}
\end{eqnarray}
with $V\to\infty$. Here, we have defined $\bar{k}\equiv-k$,
$\Delta_k=2\sqrt{2}\sin k$, and $E_{k\sigma}=U/2-(3+\sigma)\cos k$,
where $\sigma=\pm 1$ for the spin-up and spin-down dispersions,
respectively.  The infinite potential $V$ projects out the unphysical state $|\!\!\uparrow\downarrow\rangle_j$,
i.e., the state in which a doublon and a holon are on the same site.
Within the truncated local Hilbert space, this transformation is
exact.

As a further approximation, we will ignore the constraint and allow
for double occupancies by setting $V=0$, leading to the unconstrained
fermion (UF) model considered in Ref.~$ $\onlinecite{barmettler2012}. This is justified as long as the density of
doublons and holons is low, which is, at large enough interaction $U$,
the case both for the ground state and the dynamics starting from the
product state. The Hamiltonian, Eq.~\eqref{eq:HUF}, can then be
diagonalized by a Bogoliubov transformation,
\begin{align}
\label{Bogo}
c_{k\sigma} &= \gamma_{k\sigma}\cos\theta_k +i\gamma_{\bar{k}\bar{\sigma}}^\dagger \sin\theta_k,
\end{align}
with Bogoliubov angle
$\theta_k=\arctan[2\Delta_k/(E_{k\uparrow}+E_{k\downarrow})]/2$.
This leads to a diagonal Hamiltonian in the Bogoliubov quasiparticles,
$H=\sum_{k\sigma}\epsilon_{k\sigma}\gamma_{k\sigma}^\dagger\gamma_{k\sigma}$,
with dispersions
\begin{align}
\label{disp}
\epsilon_{k\sigma}&=\frac{1}{2}\sqrt{(E_{k\uparrow}+E_{k\downarrow})^2+4\Delta_k^2}+\frac{\sigma}{2}(E_{k\uparrow}-E_{k\downarrow}).
\end{align}
The ground state $|\Psi_g\rangle$ of the Bose-Hubbard model
\eqref{eq:hbhm} is then approximated by the vacuum of
the $\gamma$-quasiparticles, $\gamma_{k\sigma}|\Psi_g\rangle\equiv 0$,
while the product state $|\Psi_p\rangle$ takes the form of a BCS
state,
\begin{align}
\label{eq:bcsstate}
|\Psi_p\rangle=\prod_k\left(\cos\theta_k+i\sin\theta_k\gamma_{k\uparrow}^\dagger\gamma_{\bar{k}\downarrow}^\dagger\right)|\Psi_g\rangle.
\end{align}
Note that the doublon quasiparticle dispersion becomes gapless at
$U=8$, $\epsilon_{0\uparrow}=0$, showing that the UF approximation 
breaks down for smaller
interactions. Instead of setting $V=0$ in Eq.~\eqref{eq:HUF}, the
constraint can be taken into account approximately by a summation of
ladder diagrams which leads to a Bethe-Salpeter
equation.\cite{kotov1998} The corresponding self-energies will then
shift both quasiparticle dispersions $\epsilon_{k\sigma}$ up, so that
the model remains stable even below $U=8$. However, this formalism
cannot directly be used to calculate the nonequilibrium correlation
function we are interested in. We will therefore only use the UF model
to explain qualitative features of the defect dynamics for
interactions $U\geq 8$.

\begin{figure*}[t]
  \centering
  \includegraphics[width=2\columnwidth]{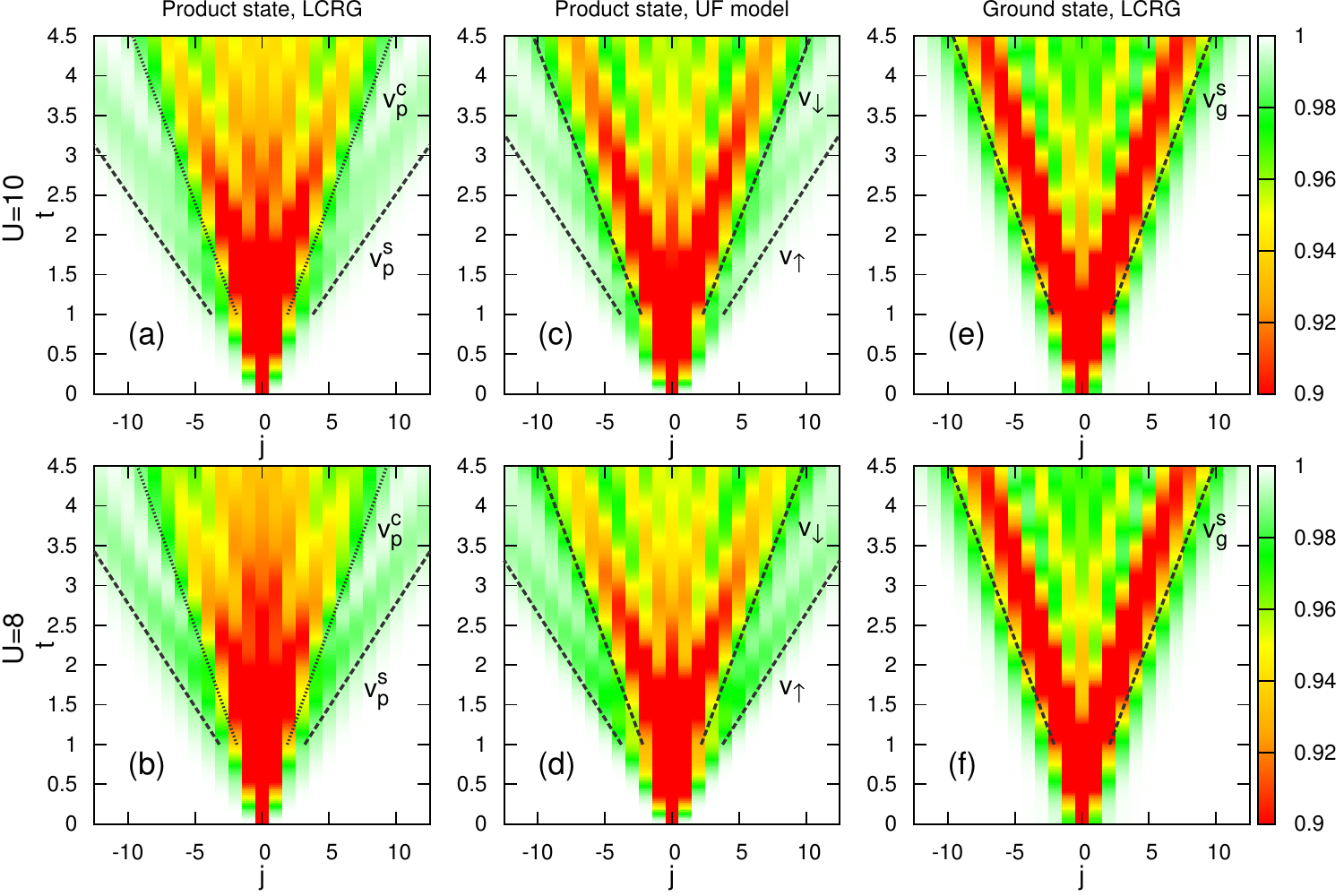}
  \caption{(color online) Density profiles $C_{p/g}(j,t)$ for large
    $U$: (a,b) Numerical results by LCRG for the initial product state
    $|\Psi_p\rangle$.  (c,d) Results for the product state by
    numerically evaluating the integrals \eqref{eq:cuf} in the UF
    model approximation. (e,f) LCRG results with the ground state
    $|\Psi_g\rangle$ as initial state. The dashed lines in (a,b) and
    (e,f) denote the numerically extracted velocities $v_p^s$ and
    $v_g^s$, respectively. The dotted lines in (a,b) are a guide to
    the eye indicating the slow core velocity $v_p^c$.  The dashed
    lines in (c,d) denote the velocities $v_\sigma$ defined in
    Eq.~\eqref{velocities}.  }
  \label{fig:densitylargeu}
\end{figure*}

\section{Results}
\label{Results}
\subsection{Large interactions}
In Fig.~\ref{fig:densitylargeu} (a) and (b), we show the numerically
obtained profiles $C_p(j,t)$ for large, but finite values of the
interaction strength. In both cases, a clear qualitative change as
compared to the $U=\infty$ case shown in Fig.~\ref{fig:bessel} is
observed. Instead of a ballistic spreading with a single velocity, we
find a separation into a central light cone spreading with velocity
$v_p^c$, and an extra contribution which moves with a faster velocity
$v_p^s$ and clearly separates itself from the main defect density. To
extract the Lieb-Robinson velocity of the first signal, $v_p^s$, from
the numerical data, we take the point $j(t)$ where the defect density
has reached half of the height of the first peak of its time-dependent
distribution as reference point, and perform a linear fit according to
$j(t)=v_p^s t$.\cite{enss2012}

Next, we show that the presence of two velocities in the defect
dynamics can be explained using the effective UF model. Expressing the
bosonic density operator in the restricted Hilbert space in terms of
the fermions, $n_j=1+c_{j\uparrow}^\dagger
c_{j\uparrow}-c_{j\downarrow}^\dagger c_{j\downarrow}$, the correlation function $C_p$ can be
calculated within the UF model approximation, see Appendix \ref{App:UF} for details.  We
obtain
\begin{widetext}
\begin{eqnarray}
\label{eq:cuf}
&&C_p(j,t) = 1-\bigg|\frac{1}{2\pi}\int_{-\pi}^\pi dk\,e^{ikj} \big( g_{k\downarrow}(t)\cos^2\theta_k+ g_{k\uparrow}(t)\sin^2\theta_k\big)\bigg|^2-\bigg|\frac{1}{2\pi}\int_{-\pi}^\pi dk\,e^{ikj}\cos\theta_k\sin\theta_k \big(g_{k\downarrow}(t)-g_{k\uparrow}(t)\big)\bigg|^2  \nonumber \\
&&\approx 1-\mathcal{J}_j^2(2t)\l(1-\frac{2j(j-2)}{U^2t^2}\r)-\frac{1}{U^2}\l[\frac{j^2}{2t^2}\mathcal{J}_j^2(4t)
-\mathcal{J}_j(2t)\left(\frac{j(j+1)}{t^2}\mathcal{J}_j(4t)+\frac{8}{t}\mathcal{J}_{j+1}(2t)-\frac{4}{t}\mathcal{J}_{j+1}(4t)\right)\r]
\end{eqnarray}
\end{widetext}
with propagators $g_{k\sigma}(t)\equiv
\langle\gamma_{k\sigma}(t)\gamma_{k\sigma}^\dagger(0)\rangle=e^{-i\epsilon_{k\sigma}t}$,
and we have performed the continuum limit. In the second line the
integrals have been evaluated within an expansion in $1/U$ up to order $U^{-2}$. The result
shows that contributions propagating
with two different velocities as well as interference terms are
present in the dynamics. The two different velocities can be determined from
the maximal slope of the two dispersion relations,
\begin{equation}
\label{velocities}
v_\sigma=\max_k\,\frac{\partial \epsilon_{k\sigma}}{\partial k}.
\end{equation}
From the $1/U$-expansion in Eq.~\eqref{eq:cuf} we see that $v_\uparrow\to
4$ and $v_\downarrow\to 2$ for $U\to\infty$. This is consistent with the discussion of
the holon and doublon dynamics at $U=\infty$ in Sec.~\ref{ss:analytics}.

\begin{figure}[ht!]
\centering
\includegraphics[width=0.99\columnwidth]{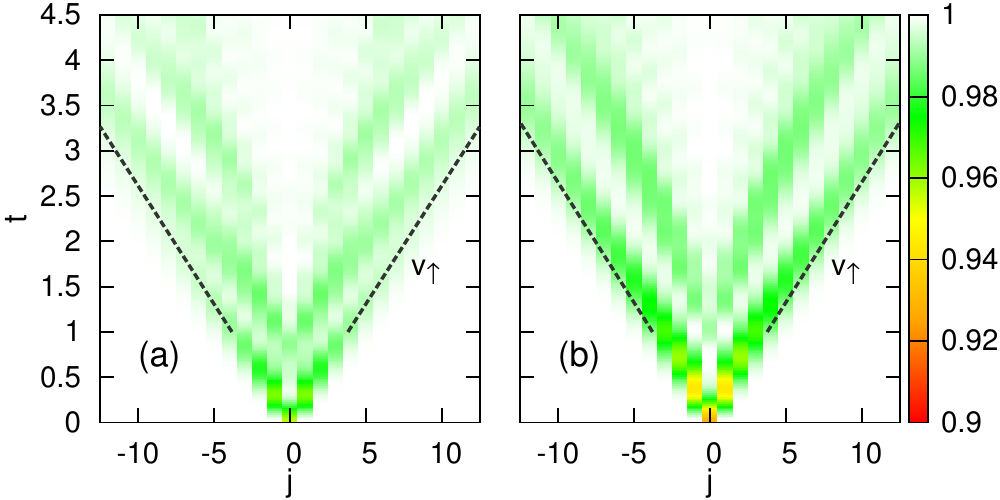}
\caption{(color online) Contributions to $C_p(j,t)$ containing the
  $g_{k\uparrow}$ propagator, see Eq.~\eqref{eq:cuf}, for (a) $U=10$ and (b)
  $U=8$.}
\label{fig:extracontrib}
\end{figure}

While the density profiles for the UF model,
Fig.~\ref{fig:densitylargeu} (c) and (d), quantitatively differ from
the LCRG results, they clearly show the same qualitative features.  In
particular, we find good agreement between the renormalized holon
velocity and the slow core velocity, $v_\downarrow\approx v_p^c$, and
the renormalized doublon velocity clearly determines the velocity of
the first signal, $v_\uparrow\approx v_p^s$. In
Fig.~\ref{fig:extracontrib} we show the contributions to
$C_p$, Eq.~\eqref{eq:cuf}, which contain the doublon propagator
$g_{k\uparrow}$, separately. Comparing with Fig.~\ref{fig:densitylargeu} (a) and
(b) we see that these contributions are indeed responsible for the
fast mode.  From the analytical result in the $1/U$-expansion, see
second line in Eq.~\eqref{eq:cuf}, we can infer the weight $W$ of these
contributions to leading order,
\begin{align}
\label{eq:weightlargeu}
W&\approx\sum_{j=-\infty}^\infty \frac{2j(j-2)}{U^2t^2}\mathcal{J}_j^2(2t)=\frac{4}{U^2}.
\end{align}

\begin{figure}[ht!]
  \centering
  \includegraphics[width=\columnwidth]{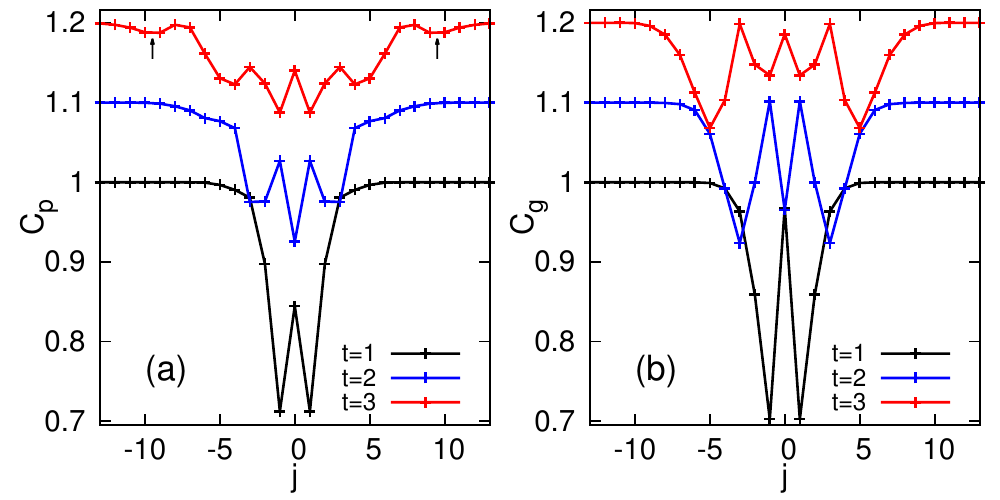}
  \caption{(color online) LCRG results for $U=8$: (a) $C_p(j,t)$
    compared to (b) $C_g(j,t)$. In (a) the position of the fast mode
    is marked by arrows. Subsequent curves are shifted by $0.1$.}
  \label{fig:profileslargeu}
\end{figure}

\begin{figure}[ht!]
\centering
\includegraphics[width=0.99\columnwidth]{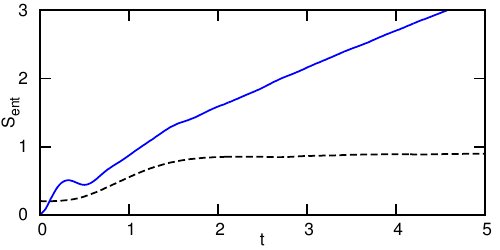}
\caption{(color online) Entanglement entropy $S_{\textrm{ent}}(t)$ for 
  $U=10$ between two halves of the infinite chain where the defect
  initially sits at the boundary. The solid line shows the linear
  increase of $S_{\textrm{ent}}(t)$ for the defect in the initial
  product state $|\Psi_p\rangle$, as expected for a global quench,
  while the dashed line represents the much slower increase of
  $S_{\textrm{ent}}(t)$ for the defect in the initial ground state
  $|\Psi_g\rangle$ (local quench).}
\label{fig:entropy}
\end{figure}

Another way to see that the mixing of holon and doublon excitations is
responsible for the fast mode is to compare with the ground state
correlation function $C_g$, shown in Fig.~\ref{fig:densitylargeu}
(e) and (f), where the fast mode is absent, and only a single velocity
$v_g^s$ can be observed. Within the UF model the initial state is now
the vacuum of the $\gamma$-quasiparticles. From
Eqs.~\eqref{eq:bosefermi} and \eqref{Bogo} one immediately sees that
in this case the $g_{k\uparrow}$ propagator does not show up in the
correlation function $C_g$; accordingly, we see that
$v_g^s\approx v_\downarrow$. The density profile therefore evolves
differently in time, as is exemplarily shown in
Fig.~\ref{fig:profileslargeu}, where cuts of the density profiles for
fixed times for the correlation functions $C_p$ and $C_g$
are compared. The difference in the defect dynamics starting from the
ground state as opposed to the product state is also clearly seen in
the entanglement entropy, $S_{\textrm{ent}}(t)$, between two halves of
the infinite chain where the defect initially sits at the boundary,
see Fig.~\ref{fig:entropy}. The extraction of a particle from the
ground state of the system constitutes a {\it local quench}. In this
case the entanglement entropy grows very slowly in time and the
numerical simulation can be performed for long
times.\cite{halimeh2014} For the initial product state, however, in
addition to the local quench, a {\it global quench} is performed,
because the initial product state $|\Psi_p\rangle$ only becomes the
exact ground state at $U=\infty$.  Hence, we see the well-known
behavior for a global quench of the entanglement entropy increasing
quickly and linearly with time.\cite{lieb1972,calabrese2005}

\begin{figure*}[t]
  \centering
  \includegraphics[width=2\columnwidth]{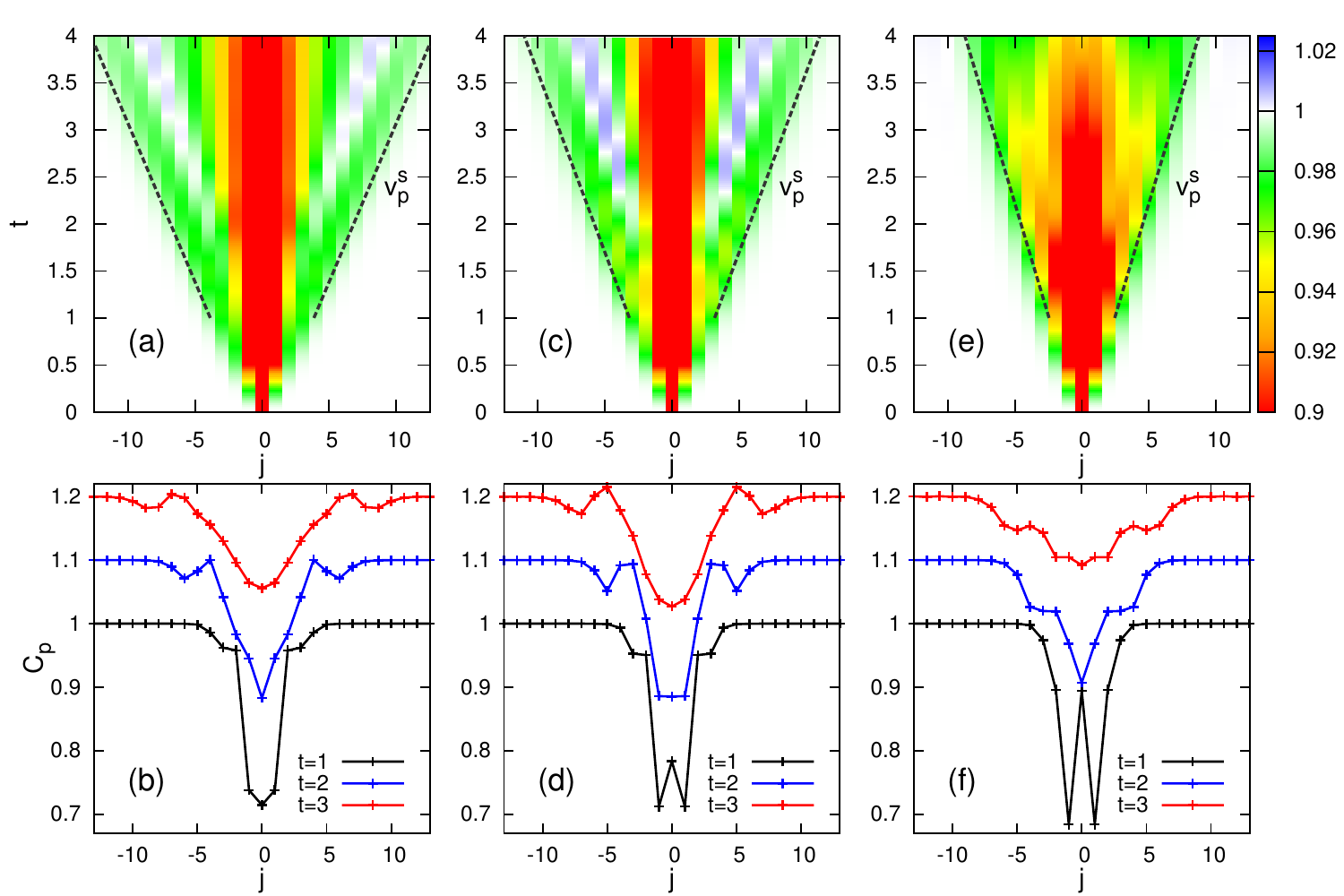}
  \caption{(color online) Density profiles $C_p(j,t)$ at weak and intermediate
    interactions for (a,b) $U=4.5$, (c,d) $U=3.0$, and
    (e,f) $U=1.5$. The dashed lines in (a,c,e) are the signal
    velocities $v_p^s$ as extracted from the numerical data. In
    (b,d,f) subsequent curves are shifted by $0.1$.}
  \label{fig:densitysmallu}
\end{figure*}

\subsection{Intermediate and weak interaction}
\begin{figure}[ht!]
  \centering
  \includegraphics[width=0.99\columnwidth]{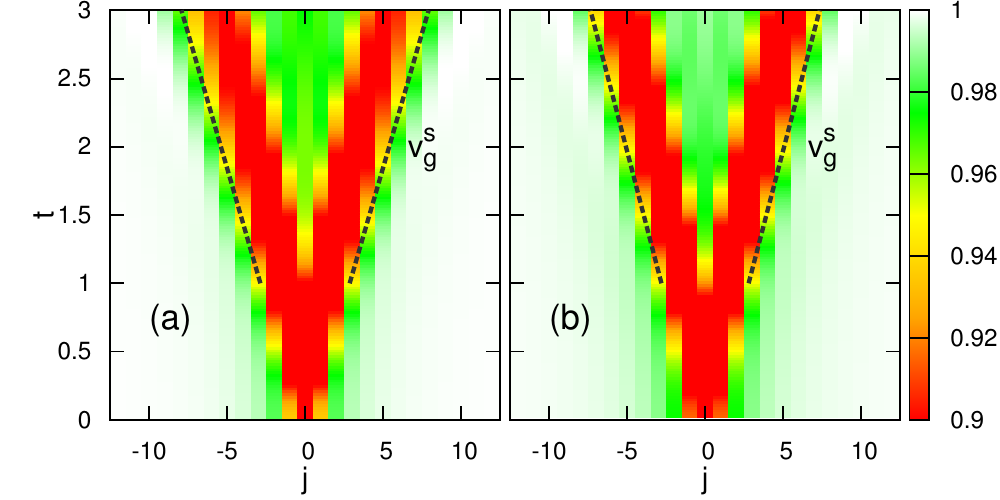}
  \caption{(Color online) Density profiles $C_g(j,t)$ starting from the ground
    state $|\Psi_g\rangle$ for (a) $U=4.5$, and (b) $U=3$. The dashed lines
    indicate the extracted signal velocities $v_g^s$.}
  \label{fig:gsrsq}
\end{figure}

For interaction strengths $U<8$ we can no longer use the UF model and
have to fully rely on numerical results. From the numerical data shown in
Fig.~\ref{fig:densitysmallu} (a-d), we see that the separation between
the fast mode and the core becomes even more pronounced for
intermediate interaction strengths.  Particularly interesting is the
case $U=3$, shown in Fig.~\ref{fig:densitysmallu} (c,d), where the
core is almost frozen in and separated from the fast mode by a region
where the density is even larger than one. While the fast mode still
spreads ballistically, we can no longer assign a velocity to the core
region, hinting at diffusive transport. The dynamics is thus
very different from that of a defect in the ground state, shown in
Fig.~\ref{fig:gsrsq}, which remains purely ballistic. For even smaller
$U$, shown in Fig.~\ref{fig:densitysmallu} (e,f), the signal velocity
$v_p^s$ decreases further so that a separation from a core region can
no longer be resolved, at least not on the numerically accessible time
scales.  Eventually, purely ballistic dynamics is restored at $U=0$,
where the density profile is again given by Eq.~\eqref{free_case}, see
Fig.~\ref{fig:bessel}.

\begin{figure}[t]
  \centering
  \includegraphics[width=0.99\columnwidth]{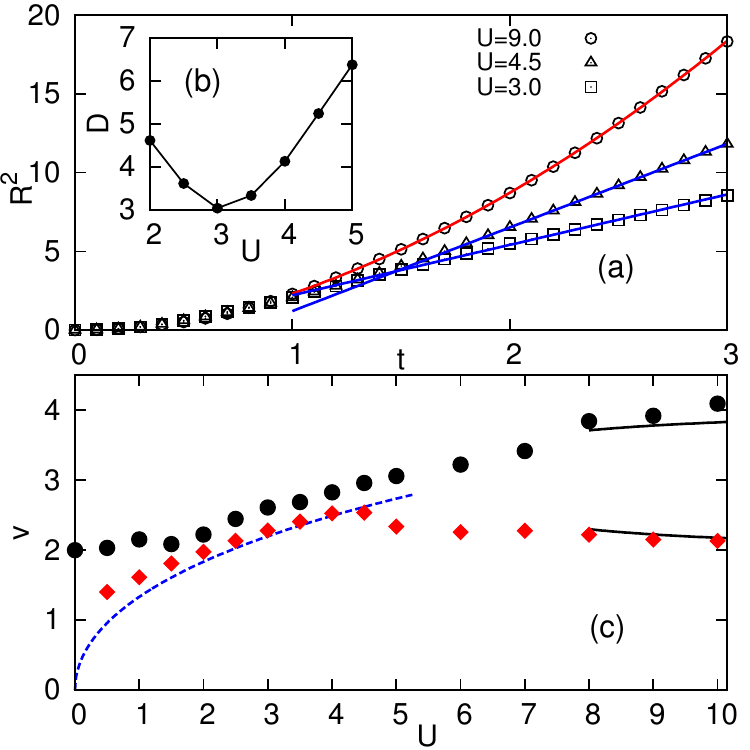}
  \caption{(color online) (a) $R^2(t)$ for the spreading of the defect
    with $|\Psi_p\rangle$ as initial state (symbols), a quadratic fit for
    $U=9$ (red line), and linear fits for $U=4.5$ and $U=3$ (blue lines).  (b) Diffusion
    constants, $R^2(t)\sim Dt$, in the regime where the dynamics is
    dominantly diffusive at intermediate times. The expansion is
    slowest close to the critical point $U_c\approx 3.37$.  (c)
    Velocity $v_p^s$ for the fastest signal in the product state
    dynamics (black circles), compared to the velocity of the signal
    $v_g^s$ for the ground state dynamics (red diamonds). For $U\geq 8$
    both velocities are well approximated by the UF model velocities
    $v_\uparrow$ and $v_\downarrow$, Eq.~\eqref{velocities},
    respectively (black solid lines).  The blue dashed line denotes
    the sound velocity $v_{\textrm{LL}}$ in the Lieb-Liniger model.}
  \label{fig:velocities}
\end{figure}

To quantify this change in the expansion dynamics of the defect for
the product state $|\Psi_p\rangle$, we calculate $R^2(t)$, as defined in
Eq.~\eqref{eq:rsq}.  The results for various values of $U$ are shown
in Fig.~\ref{fig:velocities}(a).
At short times $t\lesssim 1$ we see that the dynamics is almost
independent of the interaction strength and the defect spreads
ballistically, $R^2\sim t^2$. At longer times there is a clear
difference between large and intermediate $U$ values. While for $U=9$
the dynamics is also ballistic at intermediate times according to this
criterion, the curves for $U=4.5$ and $U=3$ become linear,
$R^2\sim D t$, indicating a dominantly diffusive spreading of the
defect at this time scale with a diffusion constant $D$.\cite{langer2009,vidmar2013} At these
intermediate timescales we can extract the diffusion constant $D$ by
linear fits; the results are shown in Fig.~\ref{fig:velocities}(b). We
see that the spreading of the core depends on interaction strength $U$
and is slowest for values close to the critical point $U_c\approx 3.37$
which separates the superfluid from the Mott insulating regime in the
ground state phase diagram. A similar slowing down near a critical
point was also observed in the free expansion dynamics of a product state into 
an empty lattice,\cite{ronzheimer2013,vidmar2013}
as well as in the non-equilibrium dynamics of the XXZ
chain.\cite{barmettler2009,barmettler2010}
Note, however, that the coexistence of diffusive and ballistic transport,
which is clearly seen in Fig.~\ref{fig:densitysmallu} (a)-(d), cannot
be resolved by the simple quantity $R^2(t)$ on intermediate timescales. In
addition, $R^2(t)$ will always be dominated by a ballistic contribution 
in the long-time limit, no matter how small this contribution is. 


A comparison of the different velocities observed in the dynamics of a
single defect is shown in Fig.~\ref{fig:velocities}(c). In contrast to
the diffusive core, critical slowing down is not observed in the
velocity of the fast ballistic mode, $v_p^s$. Instead, the velocity
$v_p^s$ is a monotonously increasing function in $U$ with $v_p^s=2$ at
$U=0$, eventually approaching the value $v_p^s\to 4$ for large $U$ as
expected for a single doublon in the initial product state, see
Sec.~\ref{ss:analytics}. For $U\geq 8$, $v_p^s$ is well approximated
by the doublon quasiparticle velocity $v_\uparrow$ determined from the
UF model. On the other hand, the velocity $v_g^s$ for the first front
of the correlation function $C_g$ is not monotonous: Here we have
$v_g^s(U=\infty)=2$, and the velocity first increases with decreasing
$U$ and approximately agrees with the holon quasiparticle velocity
$v_\downarrow$, see Eq.~\eqref{velocities}, in the regime where the UF
approximation is valid. For smaller values of $U$, $v_g^s$ decreases
with decreasing $U$. In the superfluid regime, we can then compare
$v_g^s$ with the sound velocity of the corresponding continuum
Lieb-Liniger model,\cite{calabrese2006,laeuchli2008}
\begin{equation}
\label{eq:vll}
v_{\textrm{LL}}=\sqrt{U\left(2-\sqrt{\frac{U}{2\pi^2}}\right)}.
\end{equation}
One has to keep in mind, however, that removing a particle from the
ground state is not a low energy excitation of the system, but rather
involves all momentum modes, $b_{j=0}=N^{-1} \sum_k b_k$. Only for
sufficiently large $U$ in the superfluid regime, where the
linearization of the spectrum is a good approximation for a large
region around $k=0$, can we expect that both velocities are similar.
This is indeed the case, as seen in Fig.~\ref{fig:velocities}(c).  We find,
furthermore, that the two velocities are also similar for $U$ values
even slightly above the Mott transition. Here the excitation gap is
relatively small so that the linear approximation of the dispersion
still works approximately for a large range of momenta $k>0$.

\section{Summary and Conclusions}
\label{Conclusions}
We have considered one of the simplest setups to systematically study
the non-equilibrium dynamics of the one-dimensional Bose-Hubbard
model: a single hole defect in an otherwise uniform state. If this
initial state is a product state at unit filling, we have shown
that the defect separates into a fast ballistic mode and a slower
core. This behavior, which exists for any finite interaction, can be
explained by an effective model valid for large values of $U$: doublon
quasiparticles give rise to it. Using the quantity $R^2(t)$, which
characterizes the extent of the region over which the defect has
spread, we found that the core shows signs of a diffusive time
evolution at intermediate interactions, leading to a coexistence of
ballistic and diffusive defect dynamics. While the diffusive core
shows critical slowing down, i.~e.~the dynamics is slowest close
to the quantum critical point, the velocity of the fast mode does not.
An effective theory which
describes the coexistence of a diffusive core and the ballistic
background in the dynamics and explains the slowing down of the
diffusion near the quantum critical point remains as an open problem.
We showed, furthermore, that the diffusive spreading of the defect
only occurs when starting from the product state. If one considers
instead the ground state of the Bose-Hubbard model as the initial
state, then the dynamics is always purely ballistic, and the velocity
in the superfluid regime is, for $U\gtrsim 1$, well described by the
sound velocity of the Lieb-Liniger model. 
The proposed quench protocol can easily be implemented in an ultracold
atom experiment using state-of-the-art preparation and detection
techniques with single-site resolution and offers a systematic way to
study the dynamics of local perturbations in the Bose-Hubbard model
beyond the linear response regime. Importantly, this test of the
non-equilibrium dynamics only requires to measure one-point
correlation functions, because translational invariance of the initial
state is broken by the defect.

\acknowledgments
We acknowledge support by the
Collaborative Research Centre SFB/TR49, the Graduate School of
Excellence MAINZ (DFG, Germany), as well as NSERC (Canada). We are
grateful to the Regional Computing Center at the University of
Kaiserslautern, the AHRP, and Compute Canada for providing
computational resources and support.

\appendix
\section{Density profile from the UF model}
\label{App:UF}
Our objective is to calculate the correlation function
\begin{equation}
C_p(j,t)=\langle \Psi_p|b_0^\dagger e^{iHt} n_j e^{-iHt} b_0|\Psi_p\rangle
\end{equation}
within the approximation of the UF model. Using equations \eqref{hcbosons}, \eqref{eq:bosefermi}, \eqref{Bogo}, and \eqref{eq:bcsstate}, we can express the initial state and the bosonic density operator, $n_j=1+c_{j\uparrow}^\dagger c_{j\uparrow}-c_{j\downarrow}^\dagger c_{j\downarrow}$, in terms of the $\gamma$-quasiparticles, 
\begin{widetext}
\begin{eqnarray}
b_0|\Psi_p\rangle&=&\frac{1}{\sqrt{N}}\sum_q\gamma_{\bar{q}\downarrow}^\dagger\prod_{p\not=q}\left(\cos\theta_p+i\sin\theta_p\gamma_{p\uparrow}^\dagger\gamma_{\bar{p}\downarrow}^\dagger\right)|0_\gamma\rangle,\\
n_j&=& \frac{1}{N}\sum_{k,k'} e^{i(k-k')j}\left[\cos(\theta_k-\theta_{k'})(\gamma_{k\uparrow}^\dagger\gamma_{k'\uparrow}+\gamma_{\bar{k}\downarrow}\gamma_{\bar{k}'\downarrow}^\dagger) + i\sin(\theta_k-\theta_{k'})(\gamma_{k\uparrow}^\dagger\gamma_{\bar{k}'\downarrow}^\dagger+\gamma_{\bar{k}\downarrow}\gamma_{k'\uparrow})\right]. \label{eq:densityoperator}
\end{eqnarray}
We can now calculate $C_p$ straightforwardly by using Wick's theorem. For example, the contribution of the first summand in Eq.~\eqref{eq:densityoperator} is given by
\begin{eqnarray}
&&\frac{1}{N^2}\sum_{q,k,k',q'}e^{i(k-k')j}\cos(\theta_k-\theta_{k'})\bigg\langle\gamma_{\bar{q}\downarrow}\prod_{p\not=q}(\cos\theta_p-i\sin\theta_p\gamma_{\bar{p}\downarrow}\gamma_{p\uparrow})\gamma_{k\uparrow}^\dagger(t)\gamma_{k'\uparrow}(t)\prod_{p'\not=q'}(\cos\theta_{p'}+i\sin\theta_{p'}\gamma_{p'\uparrow}^\dagger\gamma_{\bar{p}'\downarrow}^\dagger)\gamma_{\bar{q}'\downarrow}^\dagger\bigg\rangle \nonumber\\
&&= \frac{1}{N}\sum_k\sin^2\theta_k-\frac{1}{N^2}\sum_{k,k'}e^{i(k-k')j}\sin\theta_k\sin\theta_{k'}\cos(\theta_k-\theta_{k'})g_{k\uparrow}^*(t)g_{k'\uparrow}(t),
\end{eqnarray}
\end{widetext}
where we introduced the propagators $g_{k\sigma}(t)\equiv
\langle\gamma_{k\sigma}(t)\gamma_{k\sigma}^\dagger(0)\rangle$.  Here,
we can clearly see that the appearance of $g_{k\uparrow}$ propagators
is a consequence of the BCS structure of the initial product state,
which therefore are not present if the initial state is the ground
state $|0_\gamma\rangle$. Calculating the remaining three
contributions in the same way, we obtain the result given in the first
line of Eq.~\eqref{eq:cuf}.

To approximate the integrals, we can perform an
expansion in $1/U$ resulting in
\begin{eqnarray}
\cos\theta_k &\approx& 1-\frac{4}{U^2}\sin^2 k, \nonumber \\
\sin\theta_k &\approx& \frac{2\sqrt{2}}{U}\sin k+\frac{6\sqrt{2}}{U^2}\sin(2k), \nonumber \\
g_{k\sigma}(t)&\approx&e^{-itU/2}e^{-(3+\sigma)it\cos k}. 
\end{eqnarray}
Using this expansion leads to the result given in the second line of Eq.~\eqref{eq:cuf}.


\begin{thebibliography}{66}

\bibitem{kittelbook}
C.~Kittel, \emph{{Introduction to Solid State Physics}} (Wiley, 2004).

\bibitem{ashcroftbook}
N.~Ashcroft and I.~Mermin, \emph{{Solid State Physics}} (Saunders College
  Publishing, 1976), international ed.

\bibitem{greiner2002}
M.~Greiner, O.~Mandel, T.~Esslinger, T.~W. H{\"a}nsch, and I.~Bloch, Nature
  (London) \textbf{415}, 39 (2002).

\bibitem{kinoshita2006}
T.~Kinoshita, T.~Wenger, and D.~S. Weiss, Nature (London) \textbf{440}, 900
  (2006).

\bibitem{cheneau2012}
M.~Cheneau, P.~Barmettler, D.~Poletti, M.~Endres, P.~Schau{\ss}, T.~Fukuhara,
  C.~Gross, I.~Bloch, C.~Kollath, and S.~Kuhr, Nature (London) \textbf{481},
  484 (2012).

\bibitem{fukuhara2013nat}
T.~Fukuhara, P.~Schau{\ss}, M.~Endres, S.~Hild, M.~Cheneau, I.~Bloch, and
  C.~Gross, Nature (London) \textbf{502}, 76 (2013).

\bibitem{fukuhara2013nph}
T.~Fukuhara, A.~Kantian, M.~Endres, M.~Cheneau, P.~Schau{\ss}, S.~Hild,
  D.~Bellem, U.~Schollw{\"o}ck, T.~Giamarchi, C.~Gross, et~al., Nat. Phys.
  \textbf{9}, 235 (2013).

\bibitem{hild2014}
S.~Hild, T.~Fukuhara, P.~Schau{\ss}, J.~Zeiher, M.~Knap, E.~Demler, I.~Bloch,
  and C.~Gross, Phys. Rev. Lett. \textbf{113}, 147205 (2014).

\bibitem{arxivxia}
L.~Xia, L.~A. Zundel, J.~Carrasquilla, A.~Reinhard, J.~M. Wilson, M.~Rigol, and
  D.~S. Weiss (2014), arXiv:1409.2882.

\bibitem{bloch2008}
I.~Bloch, J.~Dalibard, and W.~Zwerger, Rev. Mod. Phys. \textbf{80}, 885 (2008).

\bibitem{paredes2004}
B.~Paredes, A.~Widera, V.~Murg, O.~Mandel, S.~F{\"o}lling, J.~I. Cirac, G.~V.
  Shlyapnikov, T.~W. H{\"a}nsch, and I.~Bloch, Nature (London) \textbf{429},
  277 (2004).

\bibitem{kinoshita2004}
T.~Kinoshita, T.~Wenger, and D.~S. Weiss, Science \textbf{305}, 1125 (2004).

\bibitem{gericke2008}
T.~Gericke, P.~W{\"u}rtz, D.~Reitz, T.~Langen, and H.~Ott, Nat. Phys.
  \textbf{4}, 949 (2008).

\bibitem{bakr2009}
W.~S. Bakr, J.~I. Gillen, A.~Peng, S.~F{\"o}lling, and M.~Greiner, Nature
  (London) \textbf{462}, 74 (2009).

\bibitem{wuertz2009}
P.~W{\"u}rtz, T.~Langen, T.~Gericke, A.~Koglbauer, and H.~Ott, Phys. Rev. Lett.
  \textbf{103}, 080404 (2009).

\bibitem{sherson2010}
J.~F. Sherson, C.~Weitenberg, M.~Endres, M.~Cheneau, I.~Bloch, and S.~Kuhr,
  Nature (London) \textbf{467}, 68 (2010).

\bibitem{endres2011}
M.~Endres, M.~Cheneau, T.~Fukuhara, C.~Weitenberg, P.~Schau{\ss}, C.~Gross,
  L.~Mazza, M.~C. Ba{\~n}uls, L.~Pollet, I.~Bloch, et~al., Science
  \textbf{334}, 200 (2011).

\bibitem{giamarchibook}
T.~Giamarchi, \emph{{Quantum physics in One Dimension}} (Clarendon Press,
  Oxford, 2004).

\bibitem{imambekov2012}
A.~Imambekov, T.~L. Schmidt, and L.~I. Glazman, Rev. Mod. Phys. \textbf{84},
  1253 (2012).

\bibitem{liebliniger1963}
E.~H. Lieb and W.~Liniger, Phys. Rev. \textbf{130}, 1605 (1963).

\bibitem{arxivkarrasch}
C.~Karrasch, R.~G. Pereira, and J.~Sirker (2014), arXiv:1410.2226.

\bibitem{sirker2009}
J.~Sirker, R.~G. Pereira, and I.~Affleck, Phys. Rev. Lett. \textbf{103}, 216602
  (2009).

\bibitem{sirker2011}
J.~Sirker, R.~G. Pereira, and I.~Affleck, Phys. Rev. B \textbf{83}, 035115
  (2011).

\bibitem{thurber2001}
K.~R. Thurber, A.~W. Hunt, T.~Imai, and F.~C. Chou, Phys. Rev. Lett.
  \textbf{87}, 247202 (2001).

\bibitem{degennes1958}
P.~G. {de Gennes}, J. Phys. Chem. Solids \textbf{4}, 223 (1958).

\bibitem{rigol2008}
M.~Rigol, V.~Dunjko, and M.~Olshanii, Nature (London) \textbf{452}, 854 (2008).

\bibitem{rigol2007}
M.~Rigol, V.~Dunjko, V.~Yurovsky, and M.~Olshanii, Phys. Rev. Lett.
  \textbf{98}, 050405 (2007).

\bibitem{fagotti2013}
M.~Fagotti and F.~H.~L. Essler, Phys. Rev. B \textbf{87}, 245107 (2013).

\bibitem{sirker2014}
J.~Sirker, N.~P. Konstantinidis, F.~Andraschko, and N.~Sedlmayr, Phys. Rev. A
  \textbf{89}, 042104 (2014).

\bibitem{lieb1972}
E.~H. Lieb and D.~W. Robinson, Commun. Math. Phys. \textbf{28}, 251 (1972).

\bibitem{bravyi2006}
S.~Bravyi, M.~B. Hastings, and F.~Verstraete, Phys. Rev. Lett. \textbf{97},
  050401 (2006).

\bibitem{langen2013}
T.~Langen, R.~Geiger, M.~Kuhnert, B.~Rauer, and J.~Schmiedmayer, Nat. Phys.
  \textbf{9}, 640 (2013).

\bibitem{bonnes2014}
L.~Bonnes, F.~H.~L. Essler, and A.~M. L{\"a}uchli, Phys. Rev. Lett.
  \textbf{113}, 187203 (2014).

\bibitem{kollath2005}
C.~Kollath, U.~Schollw{\"o}ck, J.~von Delft, and W.~Zwerger, Phys. Rev. A
  \textbf{71}, 053606 (2005).

\bibitem{kleine2008}
A.~Kleine, C.~Kollath, I.~P. McCulloch, T.~Giamarchi, and U.~Schollw{\"o}ck,
  New J. Phys. \textbf{10}, 045025 (2008).

\bibitem{langer2009}
S.~Langer, F.~Heidrich-Meisner, J.~Gemmer, I.~P. McCulloch, and
  U.~Schollw{\"o}ck, Phys. Rev. B \textbf{79}, 214409 (2009).

\bibitem{langer2011}
S.~Langer, M.~Heyl, I.~P. McCulloch, and F.~Heidrich-Meisner, Phys. Rev. B
  \textbf{84}, 205115 (2011).

\bibitem{karrasch2014}
C.~Karrasch, J.~E. Moore, and F.~Heidrich-Meisner, Phys. Rev. B \textbf{89},
  075139 (2014).

\bibitem{ronzheimer2013}
J.~P. Ronzheimer, M.~Schreiber, S.~Braun, S.~S. Hodgman, S.~Langer, I.~P.
  McCulloch, F.~Heidrich-Meisner, I.~Bloch, and U.~Schneider, Phys. Rev. Lett.
  \textbf{110}, 205301 (2013).

\bibitem{ganahl2012}
M.~Ganahl, E.~Rabel, F.~H.~L. Essler, and H.~G. Evertz, Phys. Rev. Lett.
  \textbf{108}, 077206 (2012).

\bibitem{alba2013}
V.~Alba, K.~Saha, and M.~Haque, J. Stat. Mech.: Theory Exp., P10018 (2013).

\bibitem{liu2014}
W.~Liu and N.~Andrei, Phys. Rev. Lett. \textbf{112}, 257204 (2014).

\bibitem{jreissaty2013}
A.~Jreissaty, J.~Carrasquilla, and M.~Rigol, Phys. Rev. A \textbf{88},
  031606(R) (2013).

\bibitem{degliespostiboschi2014}
C.~{Degli Esposti Boschi}, E.~Ercolessi, L.~Ferrari, P.~Naldesi, F.~Ortolani,
  and L.~Taddia, Phys. Rev. A \textbf{90}, 043606 (2014).

\bibitem{rodriguez2006}
K.~Rodriguez, S.~R. Manmana, M.~Rigol, R.~M. Noack, and A.~Muramatsu, New J.
  Phys. \textbf{8}, 169 (2006).

\bibitem{jreissaty2011}
M.~Jreissaty, J.~Carrasquilla, F.~A. Wolf, and M.~Rigol, Phys. Rev. A
  \textbf{84}, 043610 (2011).

\bibitem{muth2012}
D.~Muth, D.~Petrosyan, and M.~Fleischhauer, Phys. Rev. A \textbf{85}, 013615
  (2012).

\bibitem{vidmar2013}
L.~Vidmar, S.~Langer, I.~P. McCulloch, U.~Schneider, U.~Schollw{\"o}ck, and
  F.~Heidrich-Meisner, Phys. Rev. B \textbf{88}, 235117 (2013).

\bibitem{heidrichmeisner2009}
F.~Heidrich-Meisner, S.~R. Manmana, M.~Rigol, A.~Muramatsu, A.~E. Feiguin, and
  E.~Dagotto, Phys. Rev. A \textbf{80}, 041603(R) (2009).

\bibitem{kajala2011}
J.~Kajala, F.~Massel, and P.~T{\"o}rm{\"a}, Phys. Rev. Lett. \textbf{106},
  206401 (2011).

\bibitem{karlsson2011}
D.~Karlsson, C.~Verdozzi, M.~M. Odashima, and K.~Capelle, Europhys. Lett.
  \textbf{93}, 23003 (2011).

\bibitem{schneider2012}
U.~Schneider, L.~Hackerm{\"u}ller, J.~P. Ronzheimer, S.~Will, S.~Braun,
  T.~Best, I.~Bloch, E.~Demler, S.~Mandt, D.~Rasch, et~al., Nat. Phys.
  \textbf{8}, 213 (2012).

\bibitem{langer2012}
S.~Langer, M.~J.~A. Schuetz, I.~P. McCulloch, U.~Schollw{\"o}ck, and
  F.~Heidrich-Meisner, Phys. Rev. A \textbf{85}, 043618 (2012).

\bibitem{barmettler2012}
P.~Barmettler, D.~Poletti, M.~Cheneau, and C.~Kollath, Phys. Rev. A
  \textbf{85}, 053625 (2012).

\bibitem{kuehner1998}
T.~D. K{\"u}hner and H.~Monien, Phys. Rev. B \textbf{58}, 14741(R) (1998).

\bibitem{enss2012}
T.~Enss and J.~Sirker, New J. Phys. \textbf{14}, 023008 (2012).

\bibitem{white1992}
S.~R. White, Phys. Rev. Lett. \textbf{69}, 2863 (1992).

\bibitem{vidal2004}
G.~Vidal, Phys. Rev. Lett. \textbf{93}, 040502 (2004).

\bibitem{white2004}
S.~R. White and A.~E. Feiguin, Phys. Rev. Lett. \textbf{93}, 076401 (2004).

\bibitem{daley2004}
A.~J. Daley, C.~Kollath, U.~Schollw{\"o}ck, and G.~Vidal, J. Stat. Mech.:
  Theory Exp., P04005 (2004).

\bibitem{kotov1998}
V.~N. Kotov, O.~P. Sushkov, Z.~Weihong, and J.~Oitmaa, Phys. Rev. Lett.
  \textbf{80}, 5790 (1998).

\bibitem{halimeh2014}
J.~C. Halimeh, A.~W{\"o}llert, I.~P. McCulloch, U.~Schollw{\"o}ck, and
  T.~Barthel, Phys. Rev. A \textbf{89}, 063603 (2014).

\bibitem{calabrese2005}
P.~Calabrese and J.~Cardy, J. Stat. Mech.: Theory Exp., P04010 (2005).

\bibitem{barmettler2009}
P.~Barmettler, M.~Punk, V.~Gritsev, E.~Demler, and E.~Altman, Phys. Rev. Lett.
  \textbf{102}, 130603 (2009).

\bibitem{barmettler2010}
P.~Barmettler, M.~Punk, V.~Gritsev, E.~Demler, and E.~Altman, New J. Phys.
  \textbf{12}, 055017 (2010).

\bibitem{calabrese2006}
P.~Calabrese and J.~Cardy, Phys. Rev. Lett. \textbf{96}, 136801 (2006).

\bibitem{laeuchli2008}
A.~M. L{\"a}uchli and C.~Kollath, J. Stat. Mech.: Theory Exp., P05018 (2008).

\end{thebibliography}
\end{document}